\documentstyle[prl,aps,multicol,subfigure,psfig,subeqnar]{revtex}

\newcommand{\cL} { {\cal L} }

\newcommand{\cn} {\hskip 0.01in {\rm cn} \hskip 0.01in}

\newcommand{\demi}{\frac{1}{2}}

\newcommand{\be}{\begin{equation}}
\newcommand{\ee}{\end{equation}}
\newcommand{\ba}[1] {\begin{array}{ #1 }}
\newcommand{\ea}{\end{array}}

\begin{document}

\title{Stability of Attractive Bose-Einstein Condensates in a Periodic Potential}
\author{J. C. Bronski$^{1}$, L. D. Carr$^{2}$,
R. Carretero-Gonz\'alez$^{3}$
B. Deconinck$^{4}$, J. N. Kutz$^{4}$\cite{byline}, and K. Promislow$^{3}$ \\}
\address{$^{1}$Department of Mathematics, University of Illinois 
         Urbana-Champaign, Urbana, IL 61801, USA\\}
\address{$^{2}$Department of Physics, University of Washington, 
         Seattle, WA 98195-1560, USA\\}
\address{$^{3}$Department of Mathematics, Simon Fraser University,
         Burnaby, B.C., CANADA V5A 1S6\\}
\address{$^{4}$Department of Applied Mathematics, University of 
         Washington, Seattle, WA 98195-2420, USA\\}
\maketitle

\date{\today}

\begin{abstract}
  Using a standing light wave trap, a stable quasi-one-dimensional attractive
  dilute-gas Bose-Einstein condensate can be realized.  In a
  mean-field approximation, this phenomenon is modeled by the cubic nonlinear
  Schr\"odinger equation with attractive nonlinearity and an elliptic function
  potential of which a standing light wave is a special case.  New families of
  stationary solutions are presented.  Some of these solutions have neither an
  analog in the linear Schr\"odinger equation nor in the integrable nonlinear
  Schr\"odinger equation.  Their stability is examined using analytic and
  numerical methods.  Trivial-phase solutions are experimentally stable
  provided they have nodes and their density is localized in the troughs of
  the potential.  Stable time-periodic solutions are also examined. 
\end{abstract}


\begin{multicols}{2}

\section{Introduction}

Dilute--gas Bose-Einstein condensates (BECs) have been generated by many
groups using different gases which are cooled to very low temperatures and
confined in magnetic fields or standing light waves.  The sign of the atomic
coupling determines whether the interaction of the BECs is repulsive or
attractive.  Note that efficient tuning between attractive and repulsive
condensates can be achieved via a Feshbach resonance~\cite{cornish1}.
Repulsive BECs are experimentally stable~\cite{ketterle1}.  In contrast,
attractive Lithium BECs have been shown to collapse in three
dimensions~\cite{foc1,foc2,foc3}, but are predicted to be stable in
quasi-one-dimension~\cite{carr22}.  Here, we study the dynamics and stability
of quasi--one--dimensional, attractive BECs trapped in standing light waves.

The mean--field description for the macroscopic BEC wavefunction is
constructed using the Hartree--Fock approximation~\cite{hartree}, resulting in
the Gross-Pitaevskii equation~\cite{pitaevskii1,gross1}.  The
quasi-one-dimensional regime of the Gross-Pitaevskii equation 
holds when the transverse dimensions of the
condensate are on the order of its healing length and the longitudinal
dimension is much longer than its transverse dimensions~\cite{carr22,carr15}.
In this regime the BEC remains phase coherent and the governing equations are
one-dimensional.  This is in contrast to a truly 1D mean-field theory which
requires transverse dimensions on the order of or less than the atomic
interaction length~\cite{petrov1}.  In quasi--1D, the Gross-Pitaevskii equation reduces to
the cubic nonlinear Schr\"odinger equation (NLS) with a
potential~\cite{carr22,dalfovo1,key1}.

\begin{figure}[htb]
\centerline{\psfig{figure=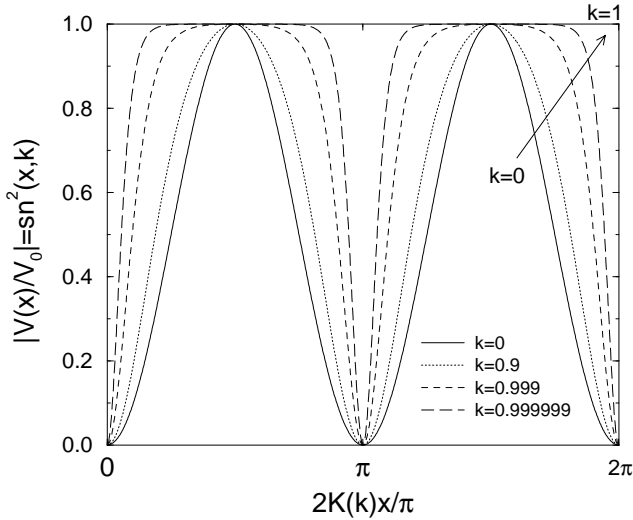,width=83mm,silent=}}
\begin{center}
\begin{minipage}{83mm}
\caption{ \label{fig:Sn_potential} The ${\rm sn}^2(x,k)$ structure of
the potential for varying
values of $k$. Note that the $x$-coordinate has been scaled by the period of the
elliptic function.  Since ${\rm sn}(x,k)$ is periodic in $x$
with period $4K(k)=4\int_0^{\pi/2}{d\alpha}/{\sqrt{1-k^2\sin^2\alpha}}$,
$V(x)$ is periodic in $x$ with period $2K(k)$.  
This period approaches infinity as $k\rightarrow 1$.} 
\end{minipage}
\end{center}
\end{figure}

In this paper we construct new solutions corresponding to a
quasi-one-dimensional attractive BEC trapped in an external periodic
potential.  The governing equation is given by the nonlinear Schr\"odinger
equation with a potential
\begin{equation}
\label{eqn:NLS}
 i\psi_t = -\frac{1}{2}\psi_{xx} - |\psi|^2 \psi 
       + V(x) \psi \, ,
\end{equation}
where $\psi(x,t)$ represents the macroscopic wave function of the condensate
and $V(x)$ is an experimentally generated macroscopic potential.  A large
class of periodic potentials is given by
\begin{equation}
    V(x) = -V_0~ {\rm sn}^2(x,k)
\label{eqn:potential}
\end{equation}
where ${\rm sn}(x,k)$ denotes the Jacobian elliptic sine function~\cite{abro}
with elliptic modulus $0\leq k\leq 1$.  In the limit $k\rightarrow 1^-$,
$V(x)$ becomes an array of well-separated hyperbolic secant potential barriers
or wells, while in the limit $k\rightarrow 0^+$ it becomes purely sinusoidal.
We note that for most intermediate values (e.g. $k=1/2$) the potential closely
resembles sinusoidal behavior and thus provides a good approximation to a
standing wave potential.  The potential is plotted in
Fig.~\ref{fig:Sn_potential} for values of $k=0, 0.9, 0.999$ and $0.999999$.
Only for $k$ very near unity (e.g. $>0.999$) does the solution start to appear
visibly elliptic. The freedom in choosing $k$ allows great flexibility in
considering a wide variety of physically realizable periodic potentials.

The paper is outlined as follows: in the next section we derive and consider
various properties and limits of two types of explicit solutions of
Eq.~(\ref{eqn:NLS}) with (\ref{eqn:potential}).  Section~III develops the
analytic framework for the linear stability properties of the new solutions of
Sec.~II.  The stability results are confirmed by numerical computations.  For
many cases, the stability analysis yields only partial analytical results, and
we rely on numerical experiments to determine stability.  Nonstationary
solutions are discussed in Sec.~IV and illustrated with various types of
time-periodic solutions.  We conclude the paper in Sec.~V with a brief summary
of the primary results of the paper and their consequences for the dynamics of
an attractive BEC.

%
%

\section{Stationary Solutions}

Equation~(\ref{eqn:NLS}) with $V(x)=0$ is an integrable equation and many
explicit solutions corresponding to various boundary conditions are known. A
comprehensive overview of these solutions is found in~\cite{belokolos}.  If
$V(x)\neq 0$, NLS is not integrable. In this case, only small classes of
explicit solutions can most likely be obtained.  Our choice of potential
~(\ref{eqn:potential}) is motivated by the form of the stationary solution of
the NLS with $V(x)=0$. An overview of these stationary solutions and their
properties is found in~\cite{carr15}.  At present, we restrict our attention to
stationary solutions of Eq.~(\ref{eqn:NLS}), $i.e.,$ solutions whose
time-dependence is restricted to  
\begin{equation} 
\psi(x,t) = r(x)~\exp(-i \omega t+i \theta(x)) \, .  
\label{eqn:ansatz} 
\end{equation} 
If $\theta_x\equiv 0$, then the solution is referred to as having trivial phase
and we choose $\theta(x)=0$.  Substituting the ansatz Eq.~(\ref{eqn:ansatz}) in
Eq.~(\ref{eqn:NLS}) and dividing out the exponential factor results in two
equations: one from the real part and one from the imaginary part. The second
equation can be integrated:  
\begin{equation} 
\theta(x) = c\int_0^x \frac{dx'}{r^2(x')} \, ,
\label{eqn:genphase} 
\end{equation} 
where $c$ is a constant of integration. Note that $\theta(x)$ is a
monotonous function of $x$. Substitution of this result in the remaining
equation gives 
\begin{equation} 
\omega r^4(x)=\frac{c^2}{2}-\frac{r^3(x)r''(x)}{2}-r^6(x)-V_0~{\rm sn}^2(x,k)
r^4(x). 
\label{eqn:ode} 
\end{equation} 
The following subsections describe two classes of solutions of this equation. 

\subsection*{Type A} 

\subsubsection{Derivation}

For these solutions, $r^2(x)$ is a quadratic function of
sn$(x,k)$: 
\begin{equation} 
r^2(x) = A~{\rm sn}^2(x,k)+B.
\label{eqn:quadratic} 
\end{equation} 
Substituting this ansatz in Eq.~(\ref{eqn:ode}) and equating the coefficients of
equal powers of ${\rm sn}(x,k)$ results in relations among the solution
parameters $\omega, c, A$ and $B$ and the equation parameters $V_0$ and $k$.
These are 
\begin{subeqnarray}\label{eqn:parametersA}
\omega &=& \frac{1}{2}\left(1+k^2-3B+\frac{BV_0}{V_0+k^2}\right),\\
c^2 &=& B~\left(\frac{B}{V_0+k^2}-1\right)\left(V_0+k^2-B k^2\right),\\
A &=& -(V_0+k^2).
\end{subeqnarray}
For a given potential $V(x)$, this solution class has one free parameter
$B$ which plays the role of a constant background level or offset. The freedom
in choosing the potential gives a total of three free parameters: $V_0$, $k$
and $B$. 

\begin{figure}[htb]
\centerline{\psfig{figure=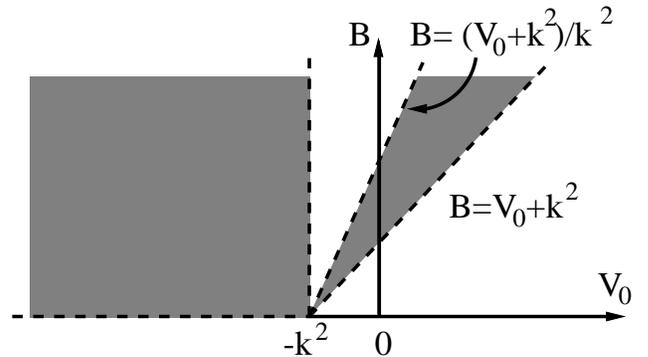,width=83mm,silent=}}
\begin{center}
\begin{minipage}{83mm}
\caption{ \label{fig:validity1} The region of validity of the solutions of Type
A is displayed shaded for a fixed value of $k$. The edges of these regions
correspond to various trivial phase solutions.}
\end{minipage}
\end{center}
\end{figure}

The requirements that both $r^2(x)$ and $c^2$ are positive imposes conditions
on the domain of these parameters:
\begin{subeqnarray}\label{eqn:validityA} 
&V_0\leq -k^2,~~ B\geq0,~~~~\mbox{or}&\\
&V_0\geq -k^2,~~ (V_0+k^2)\leq B \leq
\left(1+\displaystyle{\frac{V_0}{k^2}}\right) \, .& 
\end{subeqnarray} 
The region of validity of these solutions is displayed in
Fig.~\ref{fig:validity1}. 

%
%

%
\begin{figure}[htb]
\centerline{\psfig{figure=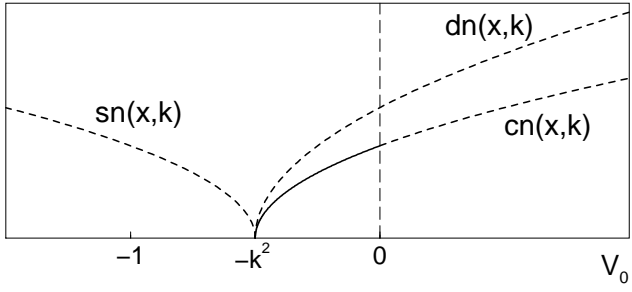,width=83mm,silent=}}
\begin{center}
\begin{minipage}{83mm}
\caption{ \label{fig:bif} The amplitude of the trivial phase solutions of Type A
versus the potential strength $V_0.$}
\end{minipage}
\end{center}
\end{figure}
%

For typical values of $V_0, k$ and $B$, the above equations give rise to
solutions of Eq.~(\ref{eqn:NLS}) which are not periodic in $x$: $r(x)$ is
periodic with period $2K(k)$, whereas $\exp(i \theta(x))$ is periodic with
period $T=\theta^{-1}(2\pi)$. In general these two periods $2K(k)$ and $T$ are
not commensurable.  Thus, requiring periodic solutions results in another condition,
namely $2K(k)/T=p/q$, for two positive integers $p$ and $q$. The most
convenient way to express this phase quantization condition is to assume the
potential ($i.e.,$ $V_0$ and $k$) is given, and to consider values of $B$ for which
the quantization condition is satisfied. Introducing $\beta=B/(V_0+k^2)$, we
find
\begin{equation}\label{eqn:genquant}
\pm\frac{\sqrt{\beta(\beta-1)(1-k^2\beta)}}{\pi}\int_0^{K(k)}\frac{dx}{-{\rm
sn}(x,k)^2+ \beta}=\frac{p}{q}. 
\end{equation}
This equation is solved for $\beta$, after which $B=\beta (V_0+k^2)$. For
numerical simulations, the number of periods of the potential is set. This
determines $q$, limiting the number of solutions of Eq.~(\ref{eqn:genquant}). 
Solutions with the same periodicity as the potential require $p/q=1$. 

Note that solutions of Type A reduce to stationary solutions of Eqs.~(\ref{eqn:NLS})
and (\ref{eqn:potential}) with $V_0=0$. Furthermore, all stationary solutions
of the integrable equation are obtained as limits of solutions of Type A.

\subsubsection{Limits and Properties}

The properties of these solutions are best understood by considering their
various limit cases. 

{\bf The trivial phase case:} The solutions of Type A have trivial phase when
$c=0$. Since $c^2$ has three factors which are linear in $B$ (see
Eq.~(\ref{eqn:parametersA})), there are three choices of $B$ for which this
occurs: $B=0$, $B=V_0+k^2$ and $B=(V_0+k^2)/k^2$. These possibilities are three
of the four boundary lines of the region of validity in
Fig.~\ref{fig:validity1}. Note that the remaining boundary line ($V_0=-k^2$)
corresponds to $r^2(x)=B$, which gives rise to a plane wave solution. Using
Jacobian elliptic function identities~\cite{abro}, one finds that the three
other boundary lines give rise to simplified solution forms: $B=0$ gives 
\begin{equation}\label{eqn:sn}
r_1(x)=\sqrt{-(V_0+k^2)}~{\rm sn}(x,k), \,\, ~\omega=\frac{1+k^2}{2}.
\end{equation}
$B=V_0+k^2$ gives
\begin{equation}\label{eqn:cn}
r_2(x)=\sqrt{V_0+k^2}~{\rm cn}(x,k), \,\, ~\omega=\frac{1}{2}-V_0-k^2,
\end{equation}
where ${\rm cn}(x,k)$ denotes the Jacobian elliptic cosine function. 
Lastly, $B=(V_0+k^2)/k^2$ gives
\begin{equation}\label{eqn:dn}
r_3(x)=\frac{\sqrt{V_0+k^2}}{k}~{\rm dn}(x,k), \,\,
~\omega=-1-\frac{V_0}{k^2}+\frac{k^2}{2},
\end{equation}
where ${\rm dn}(x,k)$ denotes the third Jacobian elliptic function.
Solution~(\ref{eqn:sn}) is valid for $V_0\leq-k^2$, whereas the other two 
solutions~(\ref{eqn:cn}) and (\ref{eqn:dn}) are valid for $V_0\geq-k^2$. 
The amplitude of
these solutions as a function of potential strength $V_0$ is shown in
Fig.~\ref{fig:bif}.

\begin{figure}[htb]
\centerline{\psfig{figure=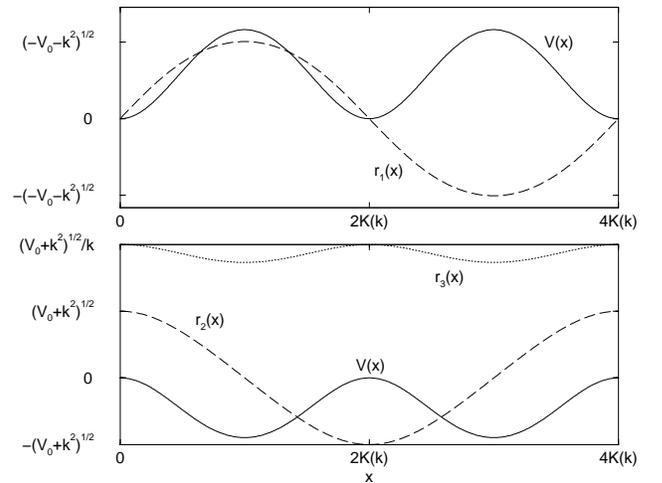,width=83mm,silent=}}
\begin{center}
\begin{minipage}{83mm}
\caption{\label{fig:triv1} Trivial phase solutions for $k=0.5$. $V(x)$ is
indicated with a solid line. For the top figure $V_0=-1$. For the bottom figure
$V_0=1$.}
\end{minipage}
\end{center}
\end{figure}

Both ${\rm cn}(x,k)$ and ${\rm sn}(x,k)$ have zero average as functions of $x$
and lie in $[-1,1]$. On the other hand, ${\rm dn}(x,k)$ has nonzero average.
Its range is $[\sqrt{1-k^2},1]$. Furthermore, ${\rm cn}(x,k)$ and ${\rm
sn}(x,k)$ are periodic in $x$ with period $4K(k)$, whereas ${\rm dn}(x,k)$ is
periodic with period $2K(k)$. 
Some solutions with trivial phase are shown in Fig.~\ref{fig:triv1}.

{\bf The trigonometric limit:} In the limit $k\rightarrow 0$, the elliptic
functions reduce to trigonometric functions and
$V(x)=-V_0~\sin^2(x)=(V_0/2)~\cos(2x)-V_0/2$. Then 
\begin{equation}\label{eqn:trig}
r^2(x)=-V_0~\sin^2(x)+B,~~~\omega=\frac{1}{2}-B.
\end{equation}
In this case, the phase integral Eq.~(\ref{eqn:genphase}) results in 
\begin{equation}\label{trig:phase}
\tan(\theta(x))=\pm\sqrt{1-V_0/B}~\tan(x).
\end{equation}
%
%
%
%
\begin{figure}[htb]
\vspace*{6mm}
\centerline{\psfig{figure=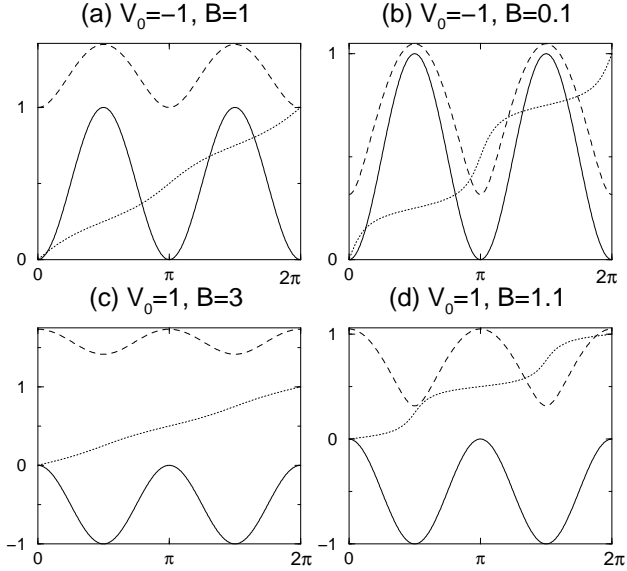,width=83mm,silent=}}
\begin{center}
\begin{minipage}{83mm}
\caption{\label{fig:trig} Phase and amplitude of the 
trigonometric solutions. For all these figures,
the solid line denotes $V(x)$, the dashed line is $r(x)$ and the dotted line is
$\theta(x)/(2\pi)$. Note that $\theta(x)$ becomes piecewise constant, as $B$
approaches the boundary of the region of validity. Far away from this boundary,
$\theta(x)$ is essentially linear.}
\end{minipage}
\end{center}
\end{figure}
%
%
\noindent
Note that this formula guarantees that the resulting solution is periodic with
the same period as the potential, so no phase quantization is required. In the
trigonometric limit, the wedge between the two regions of validity in
Fig.~\ref{fig:validity1} disappears. This is no surprise, as in this limit,
${\rm dn}(x,k)\rightarrow 1$, and the third trivial phase solution reduces
to a plane wave solution. The corner point of the region of validity also moves
to the origin. Some trigonometric solutions are illustrated in Fig.~\ref{fig:trig}.

\begin{figure}[htb]
\vspace*{6mm}
\centerline{\psfig{figure=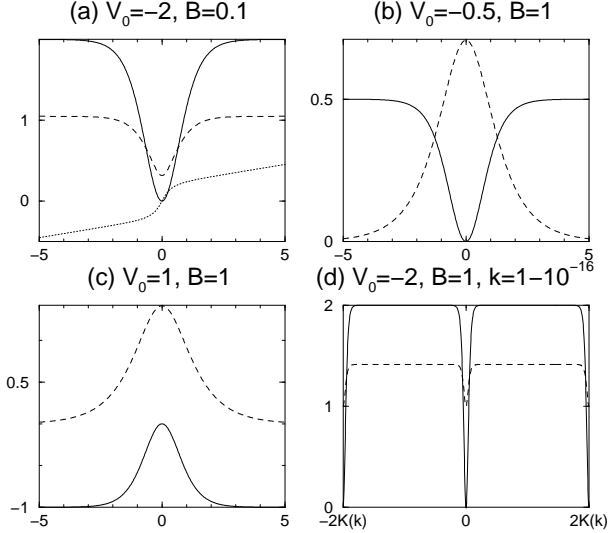,width=83mm,silent=}}
\begin{center}
\begin{minipage}{83mm}
\caption{\label{fig:solitons} Solutions with k=1 (a,b,c) or $k\rightarrow 1$
(d). The solid line denotes $V(x)$, the dashed line is
$r(x)$ and the dotted line is $\theta(x)/2\pi$. In (d), a value of $k=1-10^{-16}$
was used.}
\end{minipage}
\end{center}
\end{figure}

{\bf The solitary wave limit:} $k=1$.  In this limit the elliptic functions
reduce to hyperbolic functions. Specifically, ${\rm sn}(x,k)=
\tanh(x)$. Hence in this limit, the potential has only a
single well or a single peak. Then $V_0<0$ gives rise to a repulsive potential,
whereas $V_0>0$ gives rise to an attractive potential: $V(x)=-V_0~\tanh^2(x)$.
In this case the phase $\theta(x)$ of Eq.~(\ref{eqn:genphase}) can be calculated
explicitly: 
\begin{subeqnarray}\label{eqn:gensoliton} 
r^2(x)&=&-(V_0+1)~\tanh^2(x)+B,\\
\theta(x)\!\!&=&\!\!\sqrt{\frac{-B}{V_0+1}}x\!+\!
\arctan\left(\!\sqrt{\frac{V_0+1}{-B}}
\tanh(x)\!\right)\!,~~~~~
\end{subeqnarray} 
which is valid for $B>0$ and $V_0<-1$.  This solution is a stationary solitary
wave of depression on a positive background. It is reminiscent of the   gray
soliton solution of the NLS equation with repulsive nonlinearity. Note that
this solution exists with an attractive potential. One such solution is
illustrated in Fig.~\ref{fig:solitons}a. Another solution exists when
$B=V_0+1>0$: $r(x)=\sqrt{V_0+1}~{\rm sech}(x)$ and $\theta(x)=0$.  This
solution represents a stationary elevated solitary wave. It is a deformation of
the bright soliton solution of the NLS equation with attractive nonlinearity.
Depending on $V_0$, it exists in either an attractive ($-1<V_0<0$) or a
repulsive  ($V_0>0$) potential.  These solutions are shown in
Fig.~\ref{fig:solitons}b-c.   

Understanding the solitary wave limit facilitates the understanding of what
occurs for $k\rightarrow 1$. In this case the solutions of Type A
reduce to a periodic train of solitons with exponentially small interactions
as illustrated in Fig.~\ref{fig:solitons}d.   The exponentially small
nature of the interactions can be exploited analytically as is done
in Sec.~IV for nonstationary solutions.

\subsection*{Type B}

\subsubsection*{1.~Derivation}

For these solutions, $r^2(x)$ is linear in ${\rm sn}(x,k)$ or ${\rm dn}(x,k)$.
First we discuss the solution with ${\rm sn}(x,k)$. The quantities associated
with this solution will be denoted with a subindex 1. The quantities associated
with the ${\rm dn}(x,k)$ solution receive a subindex 2. 

Substituting
\begin{equation}\label{eqn:linear1}
r_1^2(x)=a_1~{\rm sn}(x,k)+b_1 \, , 
\end{equation}
in Eq.~(\ref{eqn:ode}) and equating different powers of ${\rm sn}(x,k)$ 
gives the relations:
\begin{subeqnarray}\label{eqn:parametersB1}
V_0&=&-\frac{3}{8}k^2 \, ,\\
\omega_1&=&\frac{1}{8}(1+k^2)-\frac{6a_1^2}{k^2} \, ,\\
c^2_1&=&-\frac{a_1^2}{4k^6}(16a_1^2-k^4)(16a_1^2-k^2) \, ,\\
b_1&=&\frac{4a_1^2}{k^2} \, .
\end{subeqnarray}
The class of potentials Eq.~(\ref{eqn:potential}) is restricted by the first of
these relations so that $V_0$ is in the narrow range
$-3k^2/8\leq V_0 \leq0$. The solution class now depends on one free amplitude
parameter $a_1$ and the free equation parameter $k$.

The region of validity of this solution is, as before, determined by the
requirements $c_1^2\geq 0$ and $r_1^2(x)\geq 0$:
\begin{equation}\label{eqn:validityB1}
\frac{k}{4}\geq|a_1|\geq\frac{k^2}{4}.
\end{equation}
%
%
%
%
\noindent
The period of $r_1(x)$ is twice the period of the potential. Requiring
periodicity in $x$ of this first solution of Type B gives 
\begin{equation}\label{eqn:genquant1}
\pm\frac{\sqrt{(16a_1^2-k^4)(k^2-16a_1^2)}}{4\pi
k^3}\!\!\int_0^{2K(k)}\!\!\!\!\!\!\!\!
\frac{dx}{\frac{4a_1}{k^2}+{\rm sn}(x,k)}
\!=\! \frac{p}{q}.
\end{equation}
For given $k$ and integers $p$, $q$, this equation is
solved for $a_1$. 

The ${\rm dn}(x,k)$ solutions are found by substituting
\begin{equation}\label{eqn:linear2}
r_2^2(x)=a_2~{\rm dn}(x,k)+b_2,\\
\end{equation}
in Eq.~(\ref{eqn:ode}). Equating different powers of ${\rm dn}(x,k)$
imposes the following constraints on the parameters:
\begin{subeqnarray}\label{eqn:parametersB2}
V_0&=&-\frac{3}{8}k^2 \, ,\\
\omega_2&=&\frac{1}{8}(1+k^2)+6a_2^2 \, ,\\
c^2_2&=&\frac{a_2^2}{4}(16a_2^2-1)(16a_2^2+k^2-1) \, ,~\\
b_2&=&-4a_2^2 \, .
\end{subeqnarray}
The class of potentials (\ref{eqn:potential}) is restricted as for the previous
solution by the first of
these relations. The solution class again depends on one free amplitude
parameter $a_2$ and the free equation parameter $k$. 

The region of validity of this solution is once more determined by the
requirements $c_2^2\geq 0$ and $r_2^2(x)\geq 0$: 
\begin{equation}\label{eqn:validityB2}
0\leq a_2 \leq\frac{\sqrt{1-k^2}}{4}.
\end{equation}

The period of $r_2(x)$ is equal to the period of the potential. Requiring
periodicity in $x$ of this second solution of Type B gives  
\begin{equation}\label{eqn:genquant2} 
\pm\frac{\sqrt{(16
a_2^2\!-\!1)(16a_2^2\!+\!k^2\!-\!1)}}{\pi} \!\! \int_0^{K(k)} \!\!\!\!\!\!\! 
 \frac{dx}{4a_2 \!-\!{\rm
dn}(x,k)}\!=\! \frac{p}{q}.
\end{equation} 
For given $k$ and integers $p$, $q$, this equation
is solved to determine $a_2$. 

In contrast to solutions of Type A, solutions of Type B do not have a
nontrivial trigonometric limit. In fact, for solutions of Type B,
this limit is identical to the limit in which the potential strength
$V_0=-3k^2/8$ approaches zero. Thus it is clear that the solutions of Type B
have no analogue in the integrable nonlinear Schr\"{o}dinger equation. However,
other interesting limits do exist. 

\subsubsection*{2.~Limits and Properties}

\begin{figure}[htb]
\centerline{\psfig{figure=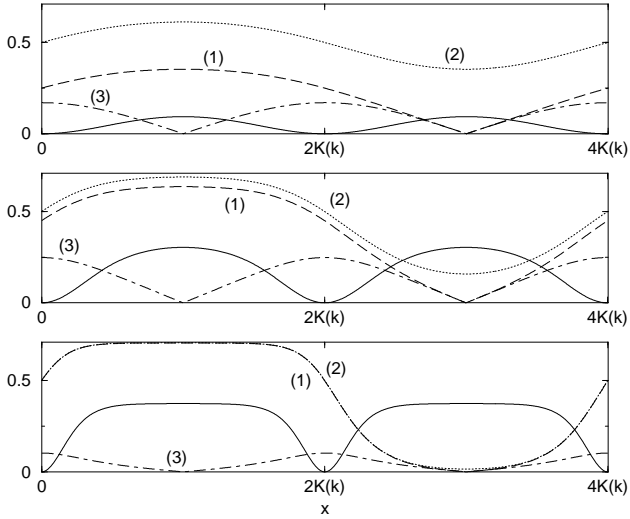,width=83mm,silent=}}
\begin{center}
\begin{minipage}{83mm}
\caption{\label{fig:trivphase2} Solutions of Type B with trivial phase. The 
figures correspond to, from top to bottom, $k=0.5$, $k=0.9$ and $k=0.999$. The
potential is indicated with a solid line. The other curves are: (1) $|r_1(x)|$
with $a_2=k^2/4$, (2) $r_1(x)$ with $a_2=k/4$ and 
(3) $|r_2(x)|$ with $a_2=k'/4$.}
\end{minipage}
\end{center}
\end{figure}

{\bf The trivial phase case:} Trivial phase corresponds to $c=0$. This occurs
precisely at the boundaries of the regions of validity. For the first solution
of Type B, there are four possibilities: $a_1=k^2/4$, $a_1=k/4$, $a_1=-k^2/4$ or
$a_1=-k/4$. By replacing
$x$ by $x+2K(k)$, one sees that the last two possibilities are completely
equivalent to the first two, so only the first two need to be considered. For
$a_1=k^2/4$, 
\begin{equation}\label{eqn:trivphasefB1}
r_1^2(x)=\frac{k^2}{4}(1+{\rm sn}(x,k))~,~~~\omega_1=\frac{1}{8}-\frac{k^2}{4}. 
\end{equation}
Equating $a_1=k/4$ gives
\begin{equation}\label{eqn:trivphasefB2}
r_1^2=\frac{1}{4}(1+k~{\rm sn}(x,k))~,~~~\omega_1=-\frac{1}{4}+\frac{1}{8}k^2.
\end{equation}
For the second solution, there are four possibilities: $a_2=0$ or
$a_2=\sqrt{1-k^2}/4$. The first 
one of these results in a zero
solution. The second one gives an interesting trivial phase solutions. 
Let $k'=\sqrt{1-k^2}$. Then, for $a_2=\sqrt{1-k^2}/4=k'/4$, 
\begin{equation}\label{eqn:trivphasefB3}
r_2^2(x)=\frac{k'}{4}({\rm dn}(x,k)-k')~,
~~~\omega_2=\frac{1}{4}+\frac{{k'}^2}{4}. 
\end{equation}
These solutions are shown in Fig.~\ref{fig:trivphase2}. 

%
%

%
\begin{figure}[htb]
\centerline{\psfig{figure=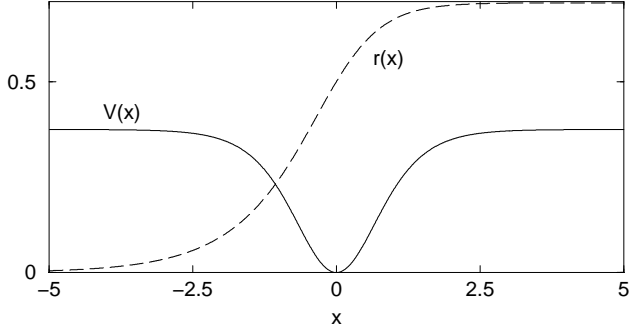,width=83mm,silent=}}
\begin{center}
\begin{minipage}{83mm}
\caption{\label{fig:solitonsB} 
Solitary wave solution of Type B. The potential is indicated with a solid
line. The dashed line solution corresponds to $a=0.3$, the dotted line to
$a=-0.3$.}
\end{minipage}
\end{center}
\end{figure}
%

{\bf The hyperbolic limit:} In this limit $V_0=-3/8$ and the potential is
$V(x)=3 \tanh(x)^2/8$. The region of validity for $r_2$ shrinks to $a_2=0$,
giving only the zero solution. However, the region of validity for $r_1$ shrinks
to $a_1=1/4$, giving rise to a non-trivial shock-like solution with trivial
phase: $r_1^2(x)=(1+\tanh(x))/4$. This solution is shown in Fig.
\ref{fig:solitonsB}. 
As for the solutions of type A, the solitary wave limit gives an idea of the
behavior of the solution for values of $k\rightarrow 1$.

\section{Stability}\label{sec:stability}

We have found a large number of new solutions to the
governing Eqs.~(\ref{eqn:NLS}) and (\ref{eqn:potential}).  However, 
only solutions that are stable can be observed in experiments.  In
this section, we consider the stability of the different solutions.
Both analytical and numerical results are presented for the solutions
with trivial phase.  In contrast, 
only numerical results are discussed for the nontrivial phase
cases.

The linear stability of the solution~(\ref{eqn:ansatz})  
is investigated.  To do
so, the exact solutions are perturbed by: 
\begin{equation}\label{eqn:perturb} 
  \psi(x,t)=(r(x)+\epsilon\phi(x,t)) \exp [{i(\theta(x)-\omega t)}]
\end{equation}
where $\epsilon \ll 1$ is a small parameter.
Collecting terms at $O(\epsilon)$ gives the linearized
equation.  Its real and imaginary parts are  
${\bf U}=(U_1,U_2)^t=(Re[\phi],Im[\phi])^t$: 
\begin{equation} 
\label{eqn:linearized}
   {\bf U}_t=JL{\bf U}=J\pmatrix{L_+ & S \cr -S & L_-} {\bf U},
\end{equation}
where 
\begin{subeqnarray}
L_+ &= & -\demi \left(\partial_x^2-\frac{c^2}{r^4} \right)-3r(x)^2 +V(x)-\omega,\\
L_- & = & -\demi \left(\partial_x^2-\frac{c^2}{r^4} \right)-r(x)^2 +V(x)-\omega,\\
S &=& \frac{c}{r(x)}\partial_x\frac{1}{r(x)},
\end{subeqnarray}
and $J=\pmatrix{0&-1\cr 1&0}$ is a skew--symmetric matrix. The operator $L$ is
Hermitian as are $L_+$ and $L_-$ while $S$ is anti--Hermitian.
Considering solutions of the form ${\bf U} (x,t) = {\hat{\bf U}}(x)\exp (\lambda t)$
gives the eigenvalue problem
\begin{equation}
\label{eqn:keith}
  \cL \hat{\bf U}=\lambda \hat{\bf U},
\end{equation}
where $\cL=JL$ and $\lambda$ is complex.  If all $\lambda$ are imaginary,
then linear stability is established.  In contrast, if there is at
least one eigenvalue with a positive real part, then instability results.
Using the phase invariance $\psi\mapsto e^{i\gamma}\psi$ of 
Eq.~(\ref{eqn:NLS}), Noether's theorem~\cite{classical} gives 
\begin{equation}
\label{eqn:null}
  \cL \pmatrix{0\cr r(x)}=0,
\end{equation}
which implies that $L_- r(x)=0.$  Thus $\lambda=0$ is in
the spectrum of $L_-$.  
For general solutions of the form (\ref{eqn:ansatz}),
determining the spectrum of the associated linearized 
eigenvalue problem (\ref{eqn:linearized}) is beyond the
scope of current methods. However, some
cases of trivial-phase solutions $(c=0)$ are amenable to analysis. 

The Hermitian operators $L_\pm$ are periodic Schr\"{o}dinger 
operators and thus the spectra of these operators is real and 
consists of bands of continuous spectrum contained in 
$[\lambda_\pm,\infty)$\cite{classical}. Here $\lambda_\pm$ denote 
the ground state eigenvalues of $L_\pm$ respectively. They are given by
\begin{equation}
  \lambda_\pm=\inf\limits_{\|\phi\|=1} \left< \phi|L_\pm |\phi 
     \right> \label{Linf} \, ,
\end{equation}
where $\|\phi\|^2=\left<\phi|\phi\right>$.   
From the relation $L_+=L_- - 2r(x)^2$ it follows that $\lambda_+ < \lambda_-$.
Also $\lambda_-\leq 0$ since $\lambda=0$ is an eigenvalue of $L_-$.

If $\lambda_- = 0$, then $L_-$ is non-negative and self-adjoint, so we can
define the non-negative square root,  $L_-^\demi$, via the spectral
theorem\cite{classical}, and hence the Hermitian operator 
$H=L_-^\demi L_+ L_-^\demi$ can be constructed. The eigenvalue 
problem for $\cL$ in Eq.~(\ref{eqn:keith}) is then equivalent to 
\begin{equation}
  (H+\lambda^2) \varphi=0,  
\end{equation}
with $\hat{U}_1=L_-^\demi \varphi$.     
Denote the left-most point of the spectrum of $H$
by $\mu_0$.  If $\mu_0\geq 0$ then $\lambda^2<0$ and
the eigenvalues of $\cL$ are imaginary and linear stability
results.   Since $H=L_-^\demi L_+
L_-^\demi$ and $L_-^\demi$ is positive, $\mu_0\geq 0$ if and only if $L_+$ is
non-negative.
In contrast, if $\mu_0<0$ then $\lambda^2>0$ and
$\cL$ has at least one pair of real eigenvalues with opposite
sign.  This shows the existence of a growing mode leading
to instability of the solution.

This non-perturbative method distinguishes between two cases
\begin{itemize}
\item[A.] If $r(x)>0$ then Eq.~(\ref{eqn:null}) implies $r(x)$  is the
ground state of $L_-$ so that $\lambda_-=0$~\cite{classical}  
while $\lambda_+<0$.  Thus the solution (\ref{eqn:ansatz}) is unstable.
\item[B.] If $r(x)$ has a zero, it is no longer the ground
state~\cite{classical} and $\lambda_-<0$.  Thus 
$\lambda_-$ and $\lambda_+$ are both negative and the
situation is indefinite.   The non-perturbative methods are insufficient
to determine linear stability or instability.  
\end{itemize}
While the solutions in Eqs.~(\ref{eqn:dn}) and (\ref{eqn:trivphasefB2}) satisfy the 
instability criterion, the stability of the remaining solutions is undetermined by
this method.  We therefore rely on 
direct computations of the nonlinear governing Eqs.~(\ref{eqn:NLS}) and
(\ref{eqn:potential}) to determine stability.  
Where the numerics indicate stability, perturbative methods are
used to confirm this observation.
For all computational simulations, twelve 
spatial periods are used.  However, to better illustrate the dynamics,
typically four spatial periods are plotted.  Moreover, all computations
are performed with white noise included in the initial data.

\subsection{trivial phase:  Type A}

\subsubsection{dn$(x,k)$}

\begin{figure}[t]
\centerline{\psfig{figure=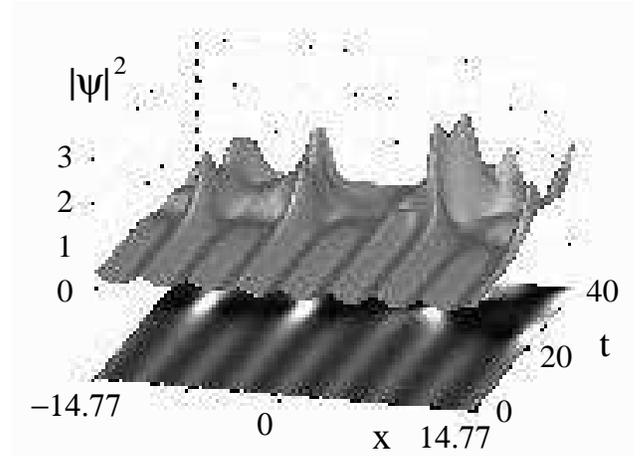,width=83mm,silent=}}
\begin{center}
\begin{minipage}{83mm}
\caption{Unstable evolution of a Type A dn$(x,k)$ solution given
by Eq.~(\ref{eqn:dn}) over 40 time units with $k=0.7$ and $V_0=-0.3$. 
\label{fig:trivialDN}}
\end{minipage}
\end{center}
\end{figure}

\begin{figure}[t]
\centerline{\psfig{figure=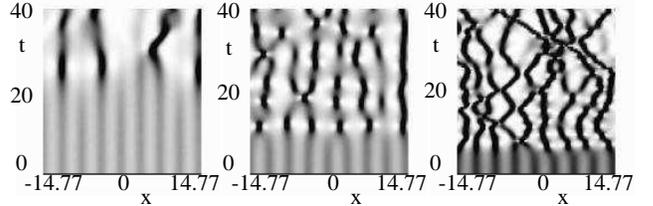,width=83mm,silent=}}
\begin{center}
\begin{minipage}{83mm}
\caption{Top view of the unstable evolution of Type A dn$(x,k)$ solutions 
given by Eq.~(\ref{eqn:dn}) over 40 time units with $k=0.7$ and with
$V_0=-0.3$ (left), $V_0=0.0$ (center), and $V_0=0.3$ (right). 
\label{fig:trivialDN2}}
\end{minipage}
\end{center}
\end{figure}

In the case of the dn$(x,k)$ solution (\ref{eqn:dn}), $r(x)>0$ and
the instability criterion A applies.   The unstable evolution
is depicted in Figs.~\ref{fig:trivialDN} and \ref{fig:trivialDN2}.
With $V_0=-0.3$ and $k=0.7$ the onset of instability occurs
at $t\approx 20$.  Figure~\ref{fig:trivialDN2} depicts an
overhead view of the dynamics with the onset
of instability for $k=0.7$ and for values
of the potential strength $V_0=-0.3, 0, 0.3$.  For increasing
values of the amplitude, $A = \sqrt{V_0+k^2}/k$, the
onset of 
%
%
%
%
%
%
%
\noindent
instability occurs earlier.   For the middle figure of 
Fig.~\ref{fig:trivialDN2}, $V_0=0$ and we are solving the integrable
nonlinear Schr\"odinger equation.  The instability of this solution
is not surprising since dn$(x,k)$ resembles a plane wave which
is known to be modulationally unstable~\cite{belokolos}.
For increasing values of the amplitude $A$, the nonlinearity
in (\ref{eqn:NLS}) becomes more dominant, resulting in an earlier 
onset of the modulational instability.

\subsubsection{sn$(x,k)$}

\begin{figure}[t]
\centerline{\psfig{figure=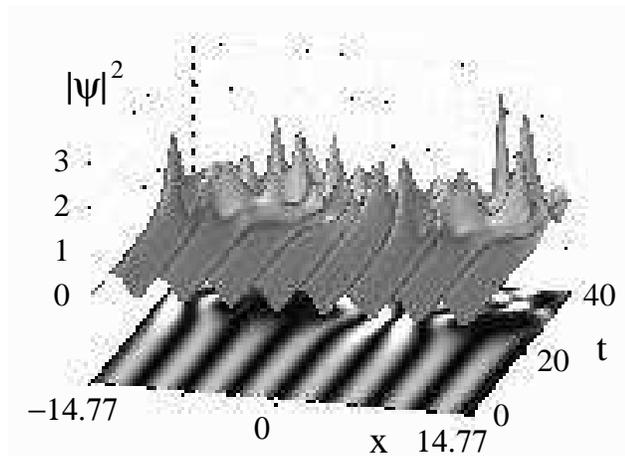,width=83mm,silent=}}
\begin{center}
\begin{minipage}{83mm}
\caption{Unstable evolution of a Type A sn$(x,k)$ solution
given by Eq.~(\ref{eqn:sn}) over 40 time units with $k=0.7$ and $V_0=-1.0$.}
\label{fig:trivialSN}
\end{minipage}
\end{center}
\end{figure}

For the sn$(x,k)$ solution (\ref{eqn:sn}), the indeterminate stability 
criterion B results.  Stability analysis for the linear Schr\"odinger equation
suggests that this case is unstable since the density is localized on
the peaks of the potential (see Fig.~\ref{fig:triv1}).   
Figure~\ref{fig:trivialSN} illustrates this behavior and shows the
onset of instability to occur for $t\approx 20$.

%
%

\subsubsection{cn$(x,k)$}

\begin{figure}[t]
\centerline{\psfig{figure=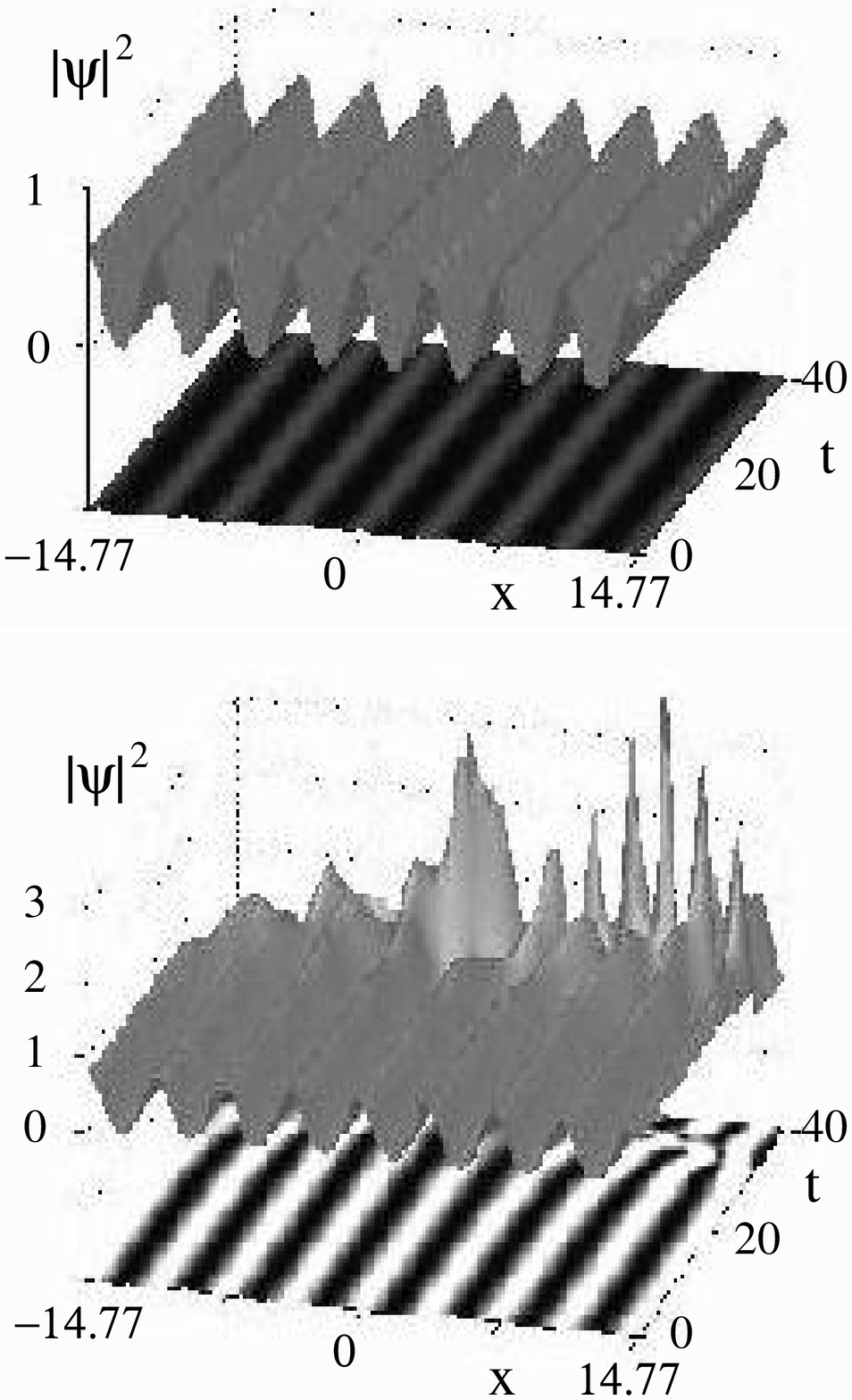,width=83mm,silent=}}
\begin{center}
\begin{minipage}{83mm}
\caption{Dynamics of Type A cn$(x,k)$ solutions given by 
Eq.~(\ref{eqn:cn}) over 40 time units with $k=0.7$ and with
$V_0=-0.3$ (top, stable) and $V_0=0.3$ (bottom, unstable).}
\label{fig:trivialCN}
\end{minipage}
\end{center}
\end{figure}

For the cn$(x,k)$ solution (\ref{eqn:cn}), the indeterminate stability 
criterion B results once again.  However, now the stability analysis 
for the linear Schr\"odinger equation
suggests that two distinct cases must be considered.  For $V_0>0$,
the density is localized on the peaks of the potential, and the solution is
unstable as illustrated in the bottom of Fig.~\ref{fig:trivialCN}.
The onset of instability occurs near $t\approx 30$.
In contrast, for $-k^2< V_0 <0$, the density is localized in
the troughs of the potential which suggests that the solution might be
stable.  Indeed, as the top of Fig.~\ref{fig:trivialCN} illustrates,
the cn$(x,k)$ solution is stable in this regime.

\subsection{Trivial Phase:  Type B}

\subsubsection{sn$(x,k)$}

\begin{figure}[t]
\centerline{\psfig{figure=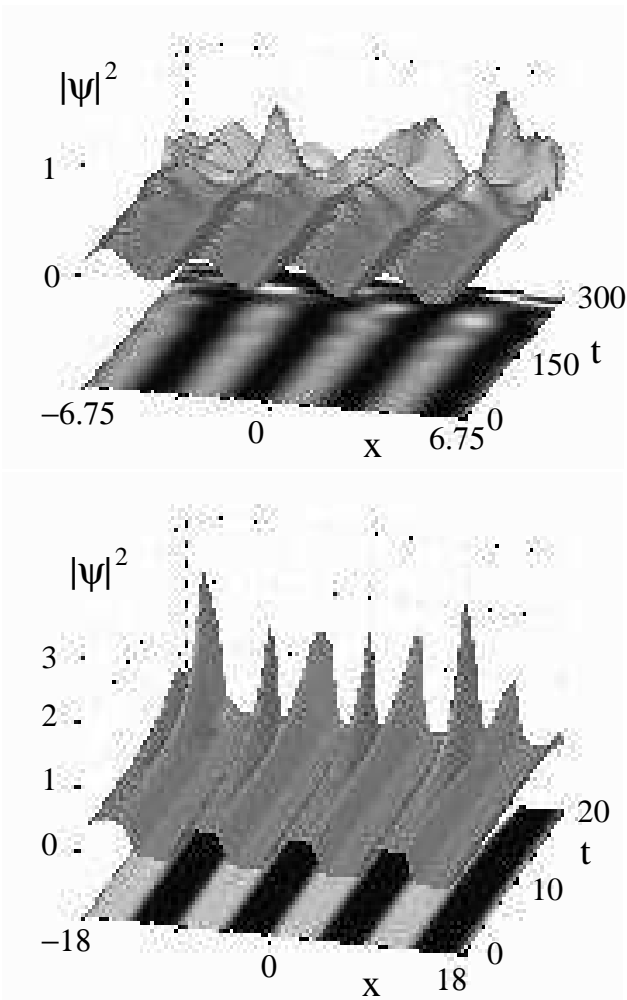,width=83mm,silent=}}
\begin{center}
\begin{minipage}{83mm}
\caption{Unstable evolution of Type B sn$(x,k)$ solutions with $a_1=k^2/4$ given
by Eq.~(\ref{eqn:trivphasefB1}) over 300 time units for $k=0.5$ (top) and 
over 20 time units for $k=0.999$. 
\label{fig:linSN1}}
\end{minipage}
\end{center}
\end{figure}

For the Type B sn$(x,k)$ solution (\ref{eqn:trivphasefB1}) with $a_1=k^2/4$,
the indeterminate stability criterion B applies.  For this solution, the
period of the density is twice the period of the potential so that the density
is not localized in the troughs of the potential.  Stability analysis for the
linear Schr\"odinger equation suggests that this case is unstable as with the
Type A sn$(x,k)$ solution.  Figure~\ref{fig:linSN1} illustrates this behavior
and shows the onset of instability to occur for $t\approx 200$ for $k=0.5$ and
$t\approx 15$ for $k=0.999$.

\begin{figure}[t]
\centerline{\psfig{figure=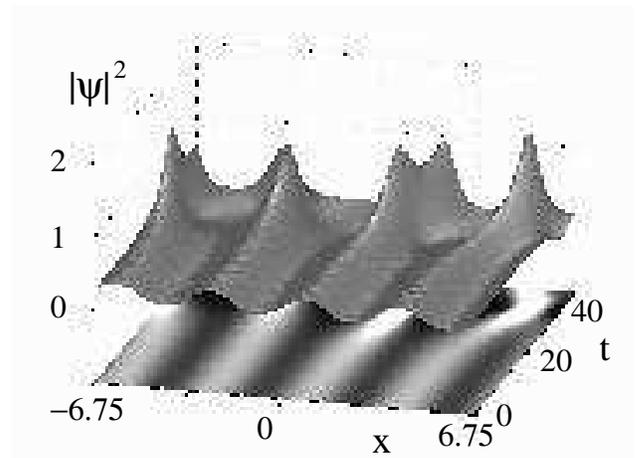,width=83mm,silent=}}
\begin{center}
\begin{minipage}{83mm}
\caption{Unstable evolution of a Type B sn$(x,k)$ solution with $a_1=k/4$ given
by Eq.~(\ref{eqn:trivphasefB2}) over 40 time units with $k=0.5$. 
\label{fig:linSN2}}
\end{minipage}
\end{center}
\end{figure}

The Type B sn$(x,k)$ solution (\ref{eqn:trivphasefB1}) with $a_1=k/4$ is
nodeless.  Thus $r(x)>0$ and the instability criterion A results.
Figure~\ref{fig:linSN2} illustrates this behavior and shows the onset of
instability to occur at $t\approx 20$ for $k=0.5$.

%
%

\subsubsection{dn$(x,k)$}

\begin{figure}[t]
\centerline{\psfig{figure=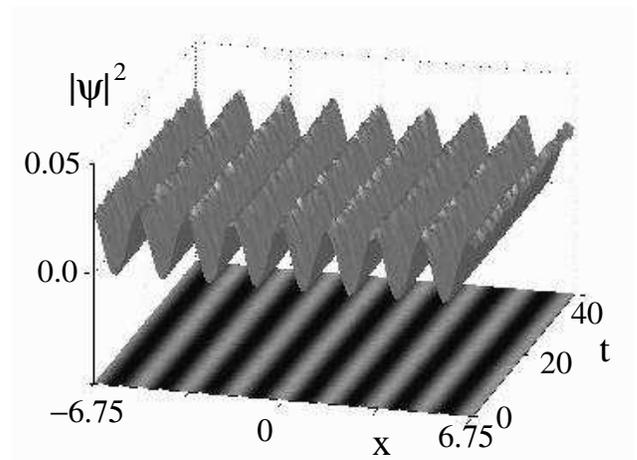,width=83mm,silent=}}
\begin{center}
\begin{minipage}{83mm}
\caption{Stable evolution of the Type B dn$(x,k)$ solution given
by Eq.~(\ref{eqn:trivphasefB3}) over 40 time units with $k=0.5$. 
\label{fig:linDN}}
\end{minipage}
\end{center}
\end{figure}

For the Type B dn$(x,k)$ solution (\ref{eqn:trivphasefB3}), the indeterminate stability 
criterion B results again.  However, the stability analysis 
for the linear Schr\"odinger equation again
suggests that the solution might be stable since the density is localized
in the troughs of the potential.  The dynamics of this solution for $k=0.5$ is
illustrated in Fig.~\ref{fig:linDN}.

\section{Nonstationary Solutions:  Spatially Extended Breathers}

Equation~(\ref{eqn:NLS}) has many solutions describing condensates that
oscillate in time.  In this section, we construct such space and time periodic
solutions in the large well separation limit ($k \rightarrow 1$).  We consider
only time-periodic solutions for which the condensate in each potential well
%
%
%
%
\noindent
oscillates with the same frequency.  More time-periodic solutions will be
considered elsewhere~\cite{breathers}.

Our solutions are obtained through a series of approximations.  First, 
we assume the ansatz
\begin{equation}\label{ansatz}
  \psi(x,t)=A \exp(-i\omega t) \sum_{k=-\infty}^\infty {\rm sech}(x-\xi_k(t)) .
\end{equation}
This equation is motivated by the fact that a trivial-phase
$\cn(x,k\rightarrow 1)$-solution can be written as $\sum_{j=-\infty}^\infty
{\rm sech}(x-4 j K(k))$ up to exponentially small terms describing the
interaction between neighboring peaks.   Note that the amplitude $A$ and frequency
$\omega$ in Eq.~(\ref{ansatz}) are fixed across the condensate.

Using the variational Lagrangian reduction approach~\cite{Perez-Garcia:96}, 
effective equations for the motion of a single lump of condensate 
are derived.  The dynamics of the center of a  lump of condensate
$\xi_k(t)$ are then given by  
%
%
%
%
%
%
%
\noindent
the
Newtonian equation of motion, 
$\ddot{\xi}_k=-\partial W(\xi_k-\xi_k^0)/\partial \xi_k$
where $\xi_k^0$ is the center position of the $k$th potential well,
with potential


%
\begin{equation}
W(\zeta) = -V_0\bar \nu \zeta^2
  \left(\alpha-\beta \zeta^2+O(\zeta^3)\right),
\end{equation}
where $\bar \nu$ is the average width of the condensate \cite{breathers},
$\alpha=4/15$ and $\beta=2/63$.  For a small displacement the condensate
undergoes near-harmonic oscillations with frequency $\sqrt{-2\alpha V_0\bar\nu}$. 

After the self-interaction, the next largest contributions are the
nearest-neighbor interactions which arise from exponentially small tail
overlaps.  These interactions result in the following lattice differential
equation (LDE) for the $k$th position $\xi_k(t)$,
%
\begin{figure}
\centerline{\psfig{file=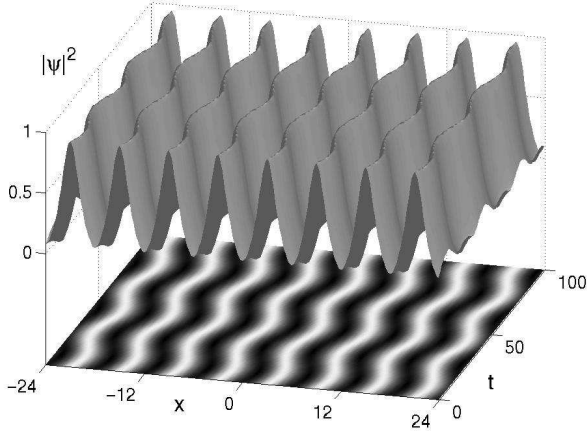,width=83mm,silent=}}
\begin{center}
\begin{minipage}{83mm}
\caption{\label{oscillations.ps} Stable vibrational mode of the condensate
corresponding to the fixed point $P_2$ over 100 time units with
$V_0=-0.1$.  The initial conditions are a
perturbation of the cnoidal wave solution (\ref{eqn:cn}) with $k=0.9791048444$
giving an initial lump-to-lump separation of $\Delta\xi_k=6$.}
\end{minipage}
\end{center}
\end{figure}
%
\begin{equation}\label{LDE}
\ddot \xi_k =
-4 A^3(e^{-A \Delta \xi_{k}}-e^{-A \Delta \xi_{k-1}})+W'(\xi_k-\xi_k^0),
\end{equation}
Here $\Delta \xi_k=\xi_{k+1}-\xi_{k}$
is the separation between centers of consecutive condensates.  The same LDE
can be derived from soliton perturbation
theory~\cite{Karpman:81a,Gerdjikov:97} or a variational
approach~\cite{Arnold:99} and corresponds to a Toda lattice~\cite{Toda:book}
with additional on-site potentials due to $W$.

We look for oscillatory solutions of Eq.~(\ref{LDE})
by considering a Fourier expansion for $\xi_k$~\cite{Bountis:00}:
\begin{equation}\label{oscillations}
   \xi_k(t) =  \sum_{j=0}^{\infty}{a_k(j) \cos(j\Omega t)}.
\end{equation}
We insert this ansatz into Eq.~(\ref{LDE}) and Taylor expand the 
exponentials, keeping terms to second order.  Equating coefficients 
of $\cos(j\Omega t)$, we obtain a recurrence relation between the amplitudes 
$a_k=a_k(j=1)$:
\begin{equation}\label{recurrence}
a_{k+1}=3\beta\,W_0\,\,a_k^3+
\left (2+4\alpha\,W_0-\bar\Omega^2\right) a_k-a_{k-1},
\end{equation}
where by using $|\Delta \xi_k^0|=2K(k)$ we find
$\bar\Omega$ = $\Omega\exp(A K(k))/2A^2$ and 
$W_0$=$(V_0/32A^3)\exp(2A K(k))$.

The recurrence relation Eq.~(\ref{recurrence}) may be written as a two 
dimensional map~\cite{ll} $(x_{k+1},y_{k+1})=F(x_k,y_k)$ by defining 
$x_k=a_{k}$ and $y_k=a_{k-1}$:
\begin{subeqnarray}\label{2Dmap}
x_{k+1} \!\!&=&\!\! 3\beta\,W_0\,\,x_k^3+
\left (2+4\alpha\,W_0-\bar\Omega^2\right) x_k-y_{k}~~~\\
y_{k+1} \!\!&=&\!\! x_k.
\end{subeqnarray}
Fixed points of this map are calculated by solving $(x_0,y_0)=F(x_0,y_0)$.
This results in three fixed points $P_i=(x_0,y_0)$ ($i=1,2,3$):  
$P_1=(0,0)$, $P_2=(x^*,y^*)$, and $P_3=(-x^*,-y^*)$ where
\begin{equation}\label{fp}
x^*=y^*=\pm\sqrt{\bar\Omega^2-4\alpha\, W_0 \over 3\beta\, W_0},
\end{equation}
%
\begin{figure}
\centerline{\psfig{file=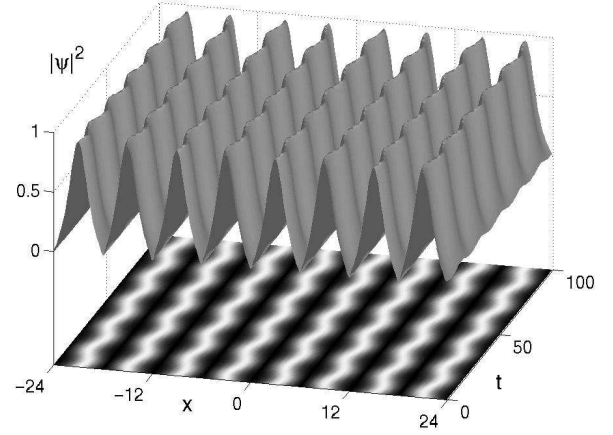,width=83mm,silent=}}
\begin{center}
\begin{minipage}{83mm}
\caption{\label{oscillations-antiphase.ps} 
Stable two-period vibrational mode of the condensate
corresponding to a period-two orbit over 100 time units with
$V_0=-0.1$.  The initial conditions are a
perturbation of the cnoidal wave solution (\ref{eqn:cn}) with $k=0.9791048444$
giving an initial lump-to-lump separation of $\Delta\xi_k=6$.
}
\end{minipage}
\end{center}
\end{figure}
%
\begin{figure}
\centerline{\psfig{file=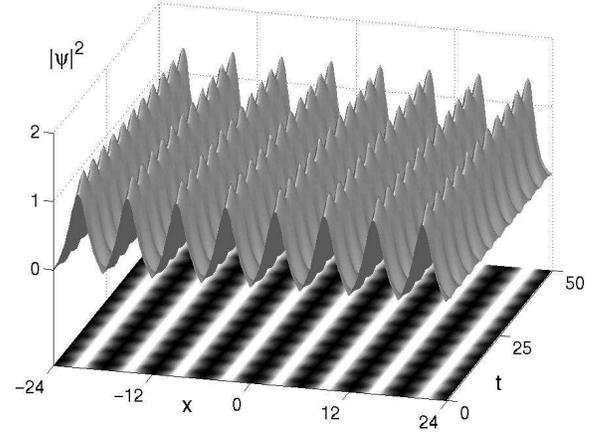,width=83mm,silent=}}
\begin{center}
\begin{minipage}{83mm}
\caption{\label{oscillations-amplitude.ps} 
Stable evolution of a breathing mode of the condensate over 50 time units with
$V_0=-0.1$.  The initial conditions are a
perturbation of the cnoidal wave solution (\ref{eqn:cn}) with $k=0.9791048444$
giving an initial lump-to-lump separation of $\Delta\xi_k=6$.
}
\end{minipage}
\end{center}
\end{figure}
%
\noindent
provided the root is real.  From Eq.~(\ref{oscillations}), 
fixed points $P_2$ and $P_3$ correspond to vibrational modes in which each
lump of condensate oscillates in time with the same amplitude, frequency $\Omega$,
and phase.  This dynamic is shown in Fig.~\ref{oscillations.ps} which
is obtained from numerical simulation of Eq.~(\ref{eqn:NLS}).


Period-two orbits of this map are calculated by solving
$(x_0,y_0)=F(F(x_0,y_0))$ and are of the form $\{(\hat{x},\hat{y}),
(-\hat{x},-\hat{y})\}$~\cite{ll}.  
This period two orbit corresponds to two alternating amplitudes  
($\hat{x}$ and $-\hat{x}$) of oscillation for consecutive condensates. 
Since the period two orbit is symmetric with respect to the origin, 
consecutive condensates oscillate with the same magnitude but 
opposite phase. The dynamics for this case are show in 
%
Fig.~\ref{oscillations-antiphase.ps}.

In a similar fashion, period-three and higher behavior can also be analyzed
with the given ansatz Eq.~(\ref{ansatz}).  In addition to allowing the lump
positions to vary in time, we can also capture amplitude time variations.  For
simplicity, we illustrate the case where only the amplitudes vary in time.
The construction of the appropriate LDE follows from previous methods.  The
fixed-point solution is illustrated in Fig.~\ref{oscillations-amplitude.ps}
where lumps of condensate oscillate in time with the same time-periodic
amplitude, frequency, and phase.  This solution type is referred to as an
extended breather.  Spatially localized breathers also exist and will be
considered elsewhere~\cite{breathers}

\section{Summary and Conclusions}

We considered the attractive nonlinear Schr\"odinger equation with an elliptic
function potential as a model for a trapped, quasi-one-dimensional
Bose-Einstein condensate.  Two new families of periodic solutions of this
equation were found and their stability was investigated both analytically and
numerically.  Additionally, stable time-periodic solutions have been analyzed.

Using perturbations with trivial phase (analysis) or perturbations with random
phase (numerics), we find that stationary trivial-phase solutions are stable
provided they have nodes and their density is localized in the troughs of the
potential.  Nodeless solutions are unstable with respect to this same class of
perturbations.  This is reminiscent of the modulational instability of the
plane wave solution of the attractive integrable nonlinear Schr\"odinger
equation.

Using random-phase perturbations, we find all nontrivial-phase solutions to be
unstable.  However, the time scale for the onset of instability for
nontrivial-phase solutions varies significantly: nontrivial-phase solutions
with parameter values close to values for trivial-phase solutions appear
unaffected by the perturbation for long times. Other nontrivial-phase solutions
go unstable more quickly.  This implies that even stable trivial-phase
solutions are structurally unstable: the smallest amount of phase ramping
causes such solutions to lose their stability, albeit on time scale which
may be longer than the lifetime of the BEC.

This result implies self-focusing of any attractive stationary condensate.
However, the effects of self-focusing can be negligible on the lifetime of the
BEC if there is no phase ramping, the density is localized in the wells of the
potential, and adjacent density peaks are separated by nodes.  Therefore, we
have demonstrated within the mean field model the existence of at least one
experimentally stable stationary state of an attractive BEC trapped in a
standing light wave.  In addition, some time-periodic condensates offer an
alternative to experimental stabilization in a standing light
wave potential.\\

{\bf Acknowledgments:} We benefited greatly from discussions with William
Reinhardt.   The work of J. Bronski, L. D. Carr, B. Deconinck, and J. N. Kutz
was supported by National Science Foundation Grants DMS--9972869, CHE97--32919,
DMS--0071568, and DMS--9802920 respectively. K. Promislow acknowledges support
from NSERC--611255

\end{multicols}

\begin{thebibliography}{10}
\bibitem[*]{byline} to whom correspondence should be addressed
\end{thebibliography}
\end{document}